\documentclass[12pt]{article}
\usepackage{a4}
\usepackage{cyrillic}
\usepackage{epsf}
\usepackage{latexsym}
\def \lsim{\,{\scriptscriptstyle{\stackrel{<}{\sim}}}\,}
\def \rsim{\,{\scriptscriptstyle{\stackrel{>}{\sim}}}\,}
\newcommand{\be}{\begin{equation}}
\newcommand{\ee}{\end{equation}}
\newcommand{\bea}{\begin{eqnarray}}
\newcommand{\eea}{\end{eqnarray}}
\newcommand{\beq}{\begin{eqnarray}}
\newcommand{\eeq}{\end{eqnarray}}
\newcommand{\beao}{\begin{eqnarray*}}
\newcommand{\eeao}{\end{eqnarray*}}
\newcommand{\nn}{\nonumber}
\newcommand{\pa}{\partial}
\newcommand{\e}{{\rm e}} 
 
\newcommand{\D}{{\rm D}} 
\newcommand{\Fc}{{\Phi_{c}}} 
\newcommand{\fc}{{\Phi_{c}}} 
\renewcommand{\i}{{\rm i}}
\renewcommand{\d}{{\rm d}}
\newcommand{\Tr}{{\rm Tr~}}

\newcommand{\Ref}[1]{(\ref{#1})}

\newcommand{\uv}{ultraviolet }

\newcommand{\la}{\lambda}

\begin{document}
\title{On Symmetry Restoration at Finite Temperature (Scalar Case)} 
\author{
{\sc M. Bordag}\thanks{e-mail: Michael.Bordag@itp.uni-leipzig.de} \\
\small  University of Leipzig, Institute for Theoretical Physics\\
\small  Augustusplatz 10/11, 04109 Leipzig, Germany\\
\small and\\
{\sc V. Skalozub}\thanks{e-mail: Skalozub@ff.dsu.dp.ua}\\
\small  Dniepropetrovsk State University, 320625 Dniepropetrovsk, Ukraine}
\maketitle
\begin{abstract}
  We investigate the effective potential for a scalar $\Phi^{4}$ theory with
  spontaneous symmetry breaking at finite temperature. All 'daisy' and 'super
  daisy' diagrams are summed up and the properties of the corresponding gap
  eqation are investigated. It is shown exactly that the phase transition is
  first order.
\end{abstract}
\section{Introduction}\label{Sec1}

The investigation of quantum fields at finite temperature is an important
problem for modern cosmology and particle physics. Since the pioneering papers
by Kirzhnits \cite{Kirzhnits} and Linde \cite{KirzhnitsLinde} numerous
investigations devoted to the development of the formalism and to various
applications appeared. The best known papers are the famous ones by Weinberg
\cite{weinberg73} and Dolan and Jackiw \cite{DolanJackiw}. There are excellent
books \cite{Linde1,Kapusta,zinnjustin,Vas2} and reviews
\cite{Linde:1979px,Gross:1981br,Mclerran:1986zb,Kalashnikov:1984sc,Braaten:1990az,Arnold:1994bp,Rubakov:1996vz}
on these topics. However, the complete understanding of the spontaneous
symmetry breaking in the early universe remains a key problem. It becomes more
important as now all particles of the Standard Model (except for the Higgs)
are discovered and their characteristics like masses are known.  Hence, a
quantitative analysis of the electroweak phase transition can be made which
allows a direct comparison with the cosmological constraints among which the
most important is the possibility of the baryogenesis scenario \cite{
  Rubakov:1996vz}, \cite{Skalozub:1999jd}. There is a lot of papers devoted to
these problems (see, for instance, \cite{LindeLinde},
\cite{Carrington:1992hz}, \cite{Arnold:1994bp} and references therein), where
the temperature effects are taken into account with different accuracy.  Of
course, in order to make such calculations reliable, the theoretical
foundations of the formalism must be deeply understood. But there are some
subtle problems to solve the present paper is devoted to.

The main problem in field theory at finite temperature with symmetry breaking
is the presence of long range correlations resp. infrared divergencies
\cite{Linde:1979px,Kalashnikov:1984sc,Gross:1981br,Arnold:1994bp}. To some
extent, the perturbative approach itself became questionable. The point is the
necessity to go beyond perturbation theory, i.e., beyond the simple expansion
in powers of the coupling constant. One has to sum up certain classes of
diagrams and to show that thereafter no infrared (or other) divergencies are
left.  There is a large number of papers trying to accomplish this task.
Quite early it was understood that the so called 'daisy' diagrams must be
summed up.  Already in \cite{DolanJackiw} it was said that even more is
necessary -- one has to consider the corresponding gap equation (an
approximated version of the Dyson equation). However, to our knowledge, up to
now no complete investigation of this kind had been done. For instance, the
problem with the imaginary part of the effective action was not solved
completely. Frequently, only the high temperature approximations of the
corresponding quantities had been used for that investigations (see for
example \cite{Takahashi:1985vx, Carrington:1992hz}).  Although the order of
magnitude is captured right by this approximation, there is a reason to expect
that a more correct calculation at finite (instead of high) temperature may
change the results for about 10 ...  20 \% which may be important in
cosmological estimates. The gap equation was considered in
\cite{Espinosa:1992gq} including a higher self energy graph in the gap
equation.  A first order transition was found, however with unwanted imaginary
parts. Similar results have been found in \cite{Nachbagauer:1995ur}.

There is another approach using flow equations based on the averaged action
method (\cite{Tetradis:1993xd} and subsequent papers) where the phase
transition was found to be of second order. Concerning the order of the phase
transition the same result was obtained in \cite{Elmfors:1992yn} using standard
renormalization group equation in $\phi_{(3)}^{6}$ and $\phi_{(4)}^{4}$
theories.  However, in both approaches there is no sufficiently strong
estimate for the neglected contributions.

In the present paper, in order to investigate these problems, we consider the
scalar $\Phi^{4}$ theory in 4(=3+1) dimensions with spontaneous symmetry
breaking at finite temperature in Matsubara representation. Using functional
methods we sum up all 'daisy' and 'super daisy' diagrams and all 'tadpole
diagrams' as well.  For the effective mass we obtain the corresponding gap
equation. All important properties of its solutions are derived exactly,
approximative methods and numerical calculations are given.  The effective
potential is shown to have a first order phase transition, all characteristic
quantities are calculated, numerically where necessary. The imaginary parts of
the effective mass and the effective potential are shown to be absent in the
minima of the effective potential at any temperature. So the summation made
turns out to deliver the essential properties of the effective potential
correctly. Further corrections may add powers in the coupling only.  The
methods used are in fact transparent and quite elementary, however some
background in functional methods is helpful.

To the authors knowledge, this result had not been obtained before. All
previous works have been done in one or other weaker approximation,
frequently using high temperature expansions.  We discuss the most
important of these approximations in detail showing their relations to
the correct values.  As it frequently happens, the correct
approximation is more elegant and technically simpler that the
intermediate ones.

In the next section the necessary functional formalism is introduced and the
summation of the infrared dangerous graphs is made. In the third section which
constitutes the main part of the paper the gap equation is investigated and
all important properties of its solution are derived. In the fourth section
the effective potential is investigated in detail and the phase transition is
shown to be of first order. In the following section the comparison with
different approaches is made. The results are discussed in the conclusions and
some basic functions used are collected in the appendix.

\section{Perturbative field theory in functional formulation}\label{Sec2}

The appropriate formulation of perturbative quantum field theory is
the functional formulation. We follow the book \cite{vasiliev} which
gives the most complete treatment of this formalism. Similar
representations are given in \cite{zinnjustin} and in \cite{fried}.

Let us first consider a general Euclidean scalar theory with the action
\be\label{a}
S(\Phi)=\frac12 \Phi K\Phi-\frac{\gamma}{3}\Phi^{3}-\frac{\la}{4}\Phi^{4}\,. 
\ee
We use condensed notations. The arguments of all quantities like the
field $\Phi$ are dropped. Wherever arguments appear they are integrated
over. The first term in the action is the 'free part', quadratic in $\Phi$.
 Its kernel is the operation $K$. In our case it is 
\be\label{k}
K=\Box -m^{2}\,.
\ee
In complete notation the action  reads 
\be\label{action} S(\Phi)=\int \d x \,\left( \frac12 \Phi(x) (\Box
-m^{2})
\Phi(x)-\frac{\gamma}{3}\Phi(x)^{3}-\frac{\la}{4}\Phi(x)^{4}\right)\,.
\ee
The generating functional of the Green functions is given by
\be\label{z} Z(J)=N \int \D \Phi \ \exp{\left(\frac12 \Phi K \Phi
-\frac{\gamma}{3}\Phi^{3}-\frac{\la}{4}\Phi^{4}+\Phi J\right)}\,, \ee
where $N$ is a normalization constant and $J$ ($J(x)$) is a source. We
consider it as a functional argument. The perturbative representation of
$Z(J)$ is
\be\label{zp} Z(J)=N \ \exp\left[\frac12{\delta\over
    \delta\Phi}\Delta{\delta\over \delta\Phi}\right] \ 
\exp{\left[-\frac{\gamma}{3}\Phi^{3}-\frac{\la}{4}\Phi^{4}+\Phi
    J\right]}_{|_{\Phi=0}}\,, \ee
where $\delta\over\delta\Phi$ denotes the variational derivative. By
expanding the exponentials the last formula turns into the
representation as the sum over all graphs with lines $\Delta =
-K^{-1}$ and generating vertices
$-\frac{\gamma}{3}\Phi^{3}-\frac{\la}{4}\Phi^{4}+\Phi J$. An actual
vertex factor is the corresponding functional derivative of the
generating vertex. A line is 'hung up' by the variational derivatives
between two (or one and the same) vertices.

The generating functional of the connected Green functions is known to be
\be\label{W}
W(J)=\ln Z(J)\,.
\ee
To pass to the effective action $\Gamma(\Phi)$ one has to perform the
Legendre transformation. For this reason one introduces the new
functional argument $\Phi={\delta\over\delta J}W(J)$, expresses $J$ in
terms of $\Phi$ and defines the effective action by
\be\label{Gamma}
\Gamma(\Phi)=W(J)-J\Phi\,.
\ee
Its perturbative expansion is known to be
\be\label{Gammapert}
\Gamma(\Phi)=S(\Phi)+\frac12 \Tr\ln\Delta+W^{{\rm 1PI}}(J\to\Delta^{-1}\Phi)\,.
\ee
Here, $S$ is the action \Ref{a}, the 'Tr ln' is the one loop
contribution and $W^{\rm 1PI}$ denotes the one particle irreducible
(1PI) graphs contributing to $W$, by means of $J\to\Delta^{-1}\Phi$
amputated for their external legs.

Now the effective potential is minus the effective action for zero
argument\footnote{Usually this notation is preserved for a constant
background field. Up to now the formulas are completely general and
the background is not specified.}:
\be\label{Veff}
V_{\rm eff}=-\Gamma(0)\,.
\ee
In the graphical representation it contains the tree contribution $S(0)$
(which is zero for the action \Ref{a}), the 'Tr ln' as the one loop
contribution and $W^{\rm 1PI}(0)$, the 1PI vacuum graphs of $W$.

In general, in the representation by graphs, lines closed over one vertex,
\epsfxsize=0.5cm\raisebox{-4pt}{\epsffile{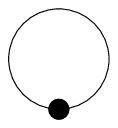}} \ , do appear.  In
field theory without external fields or without symmetry breaking resp.  at
$T=0$ these result in unessential constants and are usually hidden in a
normalization. However, for finite temperature they deliver nontrivial
contributions and must be taken seriously. In fact, from the considered
$\Phi^{4}$-vertex they appear with two external legs,
\epsfxsize=0.5cm\raisebox{-4pt}{\epsffile{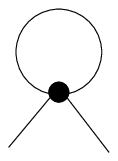}} \ .  These insertions
are called {\it daisy} graphs. They are like mass insertions. As such they
cause the known infrared problems. Consequently, they must (and, in fact, can)
be summed up completely. The same holds for {\it tadpole} insertions,
\epsfxsize=0.5cm\raisebox{-3pt}{\epsffile{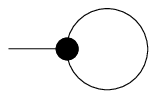}} , appearing from
$\Phi^{3}$-vertices.

The problem of summation had been tackled repeatedly. Difficulties arise
with respect to the correct count of the symmetry coefficients and beside.
Here we use a simple functional method to sum up these insertions
completely. From the point of view of the functional methods as described in
the book \cite{vasiliev} this is an easy exercise consisting in adding and
subtracting three terms in the action, i.e., in redistributing contributions
between the 'free part'  and the remaining part, which is treated
perturbatively. So we rewrite the action $S$ \Ref{a} in the form
\be\label{a2}
S=\frac12 \Phi \left(\Box-m^{2}-a^{2}\right)\Phi
-b^{2}-c \Phi +{\cal M}_{4}+{\cal M}_{3} \,,
\ee
where the new vertex factors ${\cal M}_{4}\equiv\frac{\la}{4}\Phi^{4}+\frac12
a^{2} \Phi^{2}+b^{2}$ and ${\cal M}_{3}\equiv\frac{\gamma}{3}\Phi^{3}+c\Phi$
are introduced. For the moment the parameters $a$, $b$ and $c$ are arbitrary
constants.

In the functional language all lines closed over one vertex result
from the operation of 'hanging up lines' acting only on that
vertex. So, for instance, the daisies on one $\Phi^{4}$-vertex are
generated by 
\be\label{f4} \exp\left[\frac12{\delta\over
\delta\Phi}\Delta{\delta\over \delta\Phi}\right] \ \Phi^{4} =
\Phi^{4}+6 \Delta_{0} \Phi^{2}+3 \Delta_{0}^{2}\,, \ee
where $\Delta_{0}$ is $\Delta$ for coinciding arguments (see
\Ref{D0}). A detailed explanation can be found in
\cite{vasiliev}. Nevertheless, these formulae call for a more detailed
explanation here. First of all, the line $\Delta$ has two ends and,
hence, it is the function of two arguments. For example, in free space
it is nothing else than the propagator $\Delta(x,y)=\int{\d k \over
(2\pi)^{4}}{\exp(ik(x-y))\over k^{2}+m^{2}}$. The operation with the
functional derivatives reads
\[
\frac12{\delta\over \delta\Phi}\Delta{\delta\over \delta\Phi}= \frac12 \int\d
x \int \d y \ {\delta\over\delta \Phi(x)} \Delta(x,y){\delta\over\delta
  \Phi(y)}\,.
\]
The vertex is $\Phi^{4}=\int \d x \ \Phi(x)^{4}$ and the closed line is
represented by
\be\label{D0}
\Delta_{0}=\Delta(x,y)_{|_{x=y}} = \int {\d k \over (2\pi)^{4}}{1\over
  k^{2}+m^{2}}
\ee
(for the moment we ignore the convergence problems). The index '0' is used to
symbolize that the line $\Delta(x,y)$ is taken at $x-y=0$. Now, in the
explicit notations, the derivatives can be carried out trivially and
returning back to the condensed notations the rhs. of Eq. \Ref{f4} will be
obtained.  Analogously, the relation
\be\label{f3}
\exp\left[\frac12{\delta\over \delta\Phi}\Delta{\delta\over
    \delta\Phi}\right] \ \Phi^{3} = \Phi^{3}+3 \Delta_{0} \Phi
\ee
can be checked. It describes the emergence of the tadpole diagrams from  a
$\Phi^{3}$-vertex. 

In order to avoid the appearance of 'daisies' the constants $a$ and $b$ 
introduced in Eq. \Ref{a2} may be adjusted accordingly. 
In demanding
\be\label{adj1}
\exp\left[\frac12{{\delta}\over \delta\Phi}\tilde{\Delta}{\delta\over
    \delta\Phi}\right] \ {\cal M}_{4} = -\frac{\la}{4}\Phi^{4} 
\ee
this goal can be achieved. At this point it must be noted that by
'adding zero' in Eq. \Ref{a2} we changed the first part of the action
which is the 'free part' giving rise to the lines (compare
Eqs. \Ref{z} and \Ref{zp}). Hence the mass in the propagator is
changed according to $m^{2}\to m^{2}+a^{2}$. We denote the line with
this changed mass by a tilde.

From the formula \Ref{f4} together with the simpler one
\be\label{f2}
\exp\left[\frac12{\delta\over \delta\Phi}\Delta{\delta\over
    \delta\Phi}\right] \ \Phi^{2} = \Phi^{2}+ \Delta_{0} \,,
\ee
we obtain
\beao
&\exp\left[\frac12{{\delta}\over \delta\Phi}\tilde{\Delta}{\delta\over
    \delta\Phi}\right] \exp({\cal M}_{4})
\equiv
\exp\left[\frac12{{\delta}\over \delta\Phi}\tilde{\Delta}{\delta\over
    \delta\Phi}\right] \ \left(-\frac{\la}{4}\Phi^{4}+{a^{2}\over 2}
\Phi^{2}+b^{2}\right)& \\
&=
-\frac{\la}{4}\Phi^{4}+\left(-\frac32 \la
  \tilde{\Delta}_{0}+\frac{a^{2}}{2}\right)\Phi^{2} -\frac34 \la
\tilde{\Delta}_{0}^{2}+\frac{a^{2}}{2}\tilde{\Delta}_{0}+b^{2} \,.& \nn
\eeao
In order to fulfill the requirement \Ref{adj1} the equations
\be\label{gapeq}
a^{2}=3\la \tilde{\Delta}_{0}
\ee
and
\be\label{eights}
b^{2}=-\frac34 \la \tilde{\Delta}_{0}^{2}
\ee
must be satisfied. 

In a similar way  we obtain from \Ref{f3}
\beao
&\exp\left[\frac12{{\delta}\over \delta\Phi}\tilde{\Delta}{\delta\over
    \delta\Phi}\right] \exp({\cal M}_{3})
\equiv\exp\left[\frac12{\tilde{\delta}\over \delta\Phi}\Delta{\delta\over
    \delta\Phi}\right] \ \left(-\frac{\gamma}{3}\Phi^{3}+c \Phi\right) & \\
&=-\frac{\gamma}{3}\Phi^{3}+\left(-\frac32 \gamma \tilde{\Delta}_{0}+c\right)
\Phi &
\eeao
and from the requirement
\be\label{adj2}
\exp\left[\frac12{{\delta}\over \delta\Phi}\tilde{\Delta}{\delta\over
    \delta\Phi}\right] \ {\cal M}_{3} = -\frac{\gamma}{3}\Phi^{3} 
\ee
the parameter $c$ follows to be
\be\label{lint}
c=\frac32 \gamma \tilde{\Delta}_{0}\,.
\ee
So, by virtue of \Ref{adj1} and \Ref{adj2}, no closed lines will
appear from the vertices ${\cal M}_{3}$ and ${\cal M}_{4}$. This is a
quite strong consequence because together with the 'daisy' diagrams
also all 'super daisy'\footnote{They are also called 'foam' diagrams
in \cite{Drummond:1997cw}.}  diagrams appear to be summed up. The same
holds for the 'tadpole' diagrams.  This procedure is similar to the
concept of the 'reduced vertex' in \cite{vasiliev}, p. 38.

The meaning of Eq.  \Ref{gapeq} is well known - it is a special case
of the Dyson equation, also called 'gap equation'. We shall follow the
latter terminology.  To illustrate this we denote the mass of the line
as explicit argument of the function $\tilde{\Delta}_{0}$ as
$\tilde{\Delta}_{0}(M)$, where $M^{2}=m^{2}+a^{2}$. In this notation
the $\Delta$ without tilde is ${\Delta}_{0}(m)$. So we drop the
'tilde' in what follows. Than Eq.  \Ref{gapeq} becomes
\be\label{massM}
M^{2}=m^{2}+3\la \Delta_{0}(M)  \,.
\ee
$M$ has the meaning of the {\it effective mass}.

We note  the graphical representation of $\Delta_{0}(M)$,
\be\label{delta0}
-3 \la \Delta_{0}(M) = \ \epsfxsize=0.5cm\raisebox{-4pt}{\epsffile{daisy0.eps}}
=-3\la\int{\d k \over (2\pi)^{4}}{1\over k^{2}+M^{2}} \,.
\ee

Now we turn to the corresponding theory with temperature using the Matsubara
formalism. Thereby the formulae given above remain unchanged. Further we
introduce the spontaneous symmetry breaking by giving the mass term the 'wrong
sign'.  Than, after $m^{2}\to -m^{2}$ the field $\Phi$ must be shifted
\be\label{shift}
\Phi(x) = \Fc+\eta(x)\,,
\ee
 in
order to quantize in the vicinity of the lower lying minimum of the action,
where $\Fc$ is the condensate field, a constant, and $\eta(x)$ is to be
quantized. After this substitution the action reads
\be\label{at}
S=\frac{m^{2}}{2}\Fc^{2}-\frac{\la}{4}\Fc^{4}+
\Fc (m^{2}-\la \Fc^{2})\eta +
\frac12 \eta\left(\Box-\mu^{2}\right)\eta
-\la\Fc\eta^{3}-\frac{\la}{4}\eta^{4}  \,,
\ee
where $\mu^{2}=-m^{2}+3\la \Fc^{2}$ is the new mass parameter squared. 
Using this action instead of \Ref{a} and performing the summation of the
'daisy' graphs as explained just above the action takes the form
\bea\label{atf}
S(\eta)&=&\frac{m^{2}}{2}\Fc^{2}-\frac{\la}{4}\Fc^{4}
+\frac34 \la \Delta_{0}(M)^{2}
+\Fc \left(-m^{2}+\la \Fc^{2}+3\la \Delta_{0}(M)\right)\eta \nn \\
&&+\frac12 \eta \left(\Box-M^{2}\right)\eta
-\la \Fc {\cal M}_{3}-\frac{\la}{4}{\cal M}_{4} \,,
\eea
where 
\be\label{M3}
{\cal M}_{3}=\eta^{3}+3\la \Delta_{0}(M)\eta
\ee
and 
\be\label{M4}
{\cal M}_{4}=\eta^{4}+6\la \Delta_{0}(M)\eta^{3}-\frac34 \la\Delta_{0}(M)^{2}
\ee
are the vertices which do not obtain closed loops over themselves in the
perturbative expansion assuming the equation \Ref{massM}  holds which reads
\be\label{gapteq}
M^{2}=-m^{2}+3\la\Fc^{2}+3\la \Delta_{0}(M)   
\ee
in  the present notations.
  
According to the general formula \Ref{Gammapert}, the effective action takes
now the form 
\bea\label{gat}
\Gamma(\eta)&=&-V_{\rm
  eff}+\Fc\left(-m^{2}+\la\Fc^{2}+3\la\Delta_{0}(M)\right)\eta 
+\frac12 \eta \left(\Box-M^{2}\right)\eta\nn \\
&&-\la\Fc{\cal M}_{3}-\frac{\la}{4}{\cal M}_{4}+\tilde{W}^{\rm 1PI}(\eta)  \,,
\eea
where 
\be\label{Vefft}
V_{\rm eff}=-\frac{m^{2}}{2}\Fc^{2}+\frac{\la}{4}\Fc^{4}
-\frac34\la\Delta_{0}^{2}(M)+V_{1}(M)-W^{\rm 1PI}(0) 
\ee
is the effective potential and 
\[
V_{1}= - \frac12 \ \Tr \ln \Delta
\]
is the one loop contribution with the mass $M$ in the propagator, see
Eq. \Ref{A1}. 

In the effective action $\Gamma(\eta)$, \Ref{gat}, as well as in the
action \Ref{atf}, the term linear in $\eta$ acquired an additional
contribution.  The effective mass $M$ in the addendum quadratic in
$\eta$ must obey the gap equation \Ref{gapteq} and becomes a function
of the condensate $\Fc$ and of the temperature in this way:
$M(\Fc;\la,T)$.  All higher graphs with external legs are contained in
$\tilde{W}^{\rm 1PI}$.

In the effective potential $V_{\rm eff}$ the first two terms are the
so called 'tree' approximation. The corresponding minimum of the
effective potential is reached at $\Fc=m/\sqrt{\la}$ which is at once
the condition for the linear term in $\Gamma$ to vanish at tree level
and the mass of the field is $M^{2}=2m^{2}$, then.

The third term in $V_{\rm eff}$ can be represented graphically as
\epsfxsize=0.5cm\raisebox{-8pt}{\epsffile{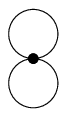}} looking like a
'eight'.  The function $V_{1}$ is a notation for the 'Tr ln'
contribution and $W^{\rm 1PI}(0)$ are the higher loop graphs without
external legs (vacuum graphs). We do not need to consider them in this
paper.

The effective potential has to be considered together with the gap equation
\Ref{gapteq}. This allows to exclude $\Delta_{0}(M)$ from explicitely
appearing  in the expression
\be\label{Veff2}
V_{\rm eff}= -{\left(M^{2}+m^{2}\right)^{2}\over 12 \la}+{M^{2}\over
  2}\Fc^{2}-\frac{\la}{2}\Fc^{4}+V_{1}(M) \,,
\ee
which, together with \Ref{gapteq}, will be used in the following.

The functions $\Delta_{0}(M)$ and $V_{1}(M)$ can be written down quite
explicitely. This had been done in a number of papers. Different
representations are possible. We collect the two most useful ones in the
Appendix together with their expansions for high $T$ which is at once the
expansion for small $M$. These quantities contain \uv divergencies which are
removed by a standard procedure which includes also the fixing of the
normalization conditions.  They are chosen not to change the minimum of the
effective potential known at tree level in case of zero temperature (the
finite temperature contributions are \uv finite and unique). So we require
\[
{\pa V_{1}\over\pa\Fc}_{|\Fc=m/\sqrt{\la}}=0 \quad \mbox{and}\quad {\pa^{2}
  V_{1}\over\pa\Fc^{2}}_{|\Fc=m/\sqrt{\la}}=0\,. 
\]
In fact, the superficial divergence degree of vacuum graphs is four. Because
$V_{\rm eff}$ depends on $\Fc^{2}$ rather than on $\Fc$, there are only three
independent normalization conditions to impose. After the first two there
remains one condition fixating the value of $V_{\rm eff}$ at some value of
$\Fc^{2}$. It is natural to demand $V_{\rm eff}(\Fc=0)=0$. But the calculation
of the corresponding constant to be subtracted from $V_{\rm eff}$ is not easy.
So we leave that condition open.  In fact, it becomes important only in figure
\ref{figure4} where it had been adjusted numerically.

The single 'daisy' graph $\Delta_{0}$, which is in fact the first graph in the
self energy, has superficial degree two and two normalization conditions have
to be imposed. In the same aim as above we choose
\[
\Delta(M^{2})_{|\Fc=m/\sqrt{\la}}=0 \,.
\]
The second condition results in an additive constant. We choose it so
that the equation
\be\label{zh}
{1\over M}{\pa V_{1}(M)\over\pa  M}= \Delta_{0}(M)\,
\ee
holds.  Note that this relation holds for the temperature dependent part
automatically.
\section{Solving the Gap Equation}\label{Sec3}

The solution of the gap equation is the key element of the whole procedure.
There is, of course, no explicit solution. However, the most important
properties can be derived exactly.

\begin{figure}[!htbp]\unitlength=1cm
\begin{picture}(5,4.5)
\put(0,0){\epsfxsize=7cm 
\epsffile{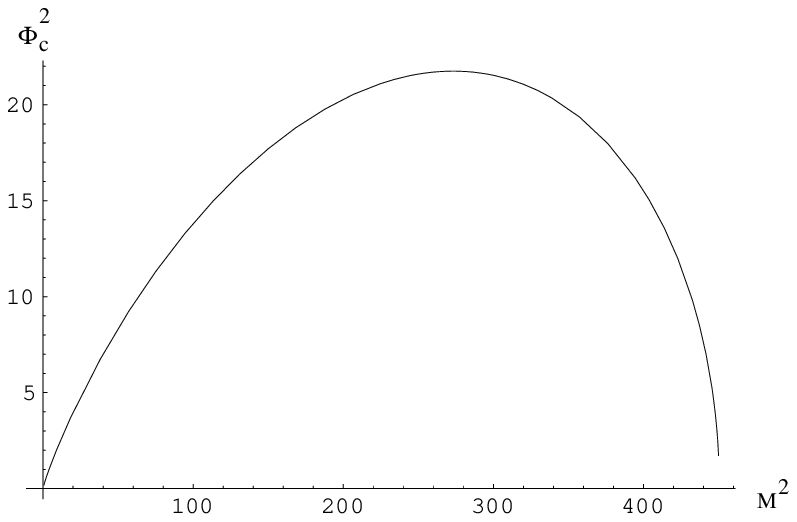}}
\end{picture}
\caption{The function $\Fc^{2}(M^{2})$ as given by the gap equation 
\Ref{gapteq} for
  $\la=5$ and $T=1$}
\label{figure1}
\end{figure}

\begin{figure}[!htbp]\unitlength=1cm
\begin{picture}(10,4.5)
\put(0,0){\epsfxsize=7cm\epsffile{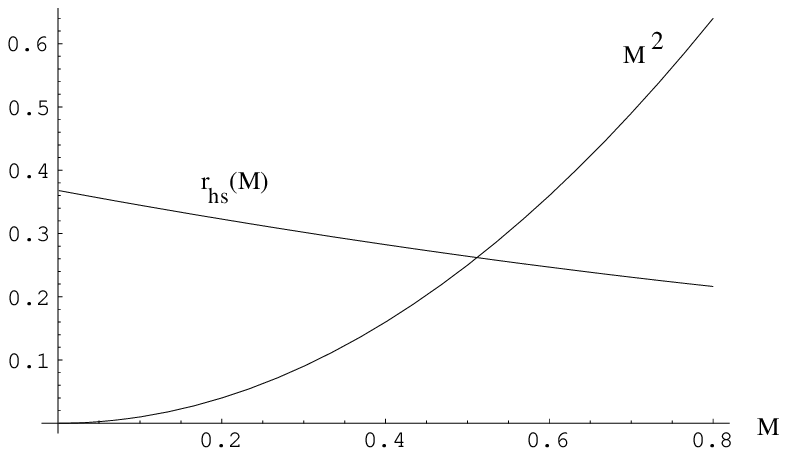}}
\end{picture}
\caption{Plot of the rhs. and the lhs. of the gap equation \Ref{gapteq} for
  $\Fc=0.6$, $\la=1$, $T=1$}
\label{figure2}
\end{figure}

In solving the gap equation \Ref{gapteq} we obtain the effective mass $M$
appearing after summing up the 'daisy' diagrams. It is a function of the
condensate $\Fc$, further it depends on the bare mass $m$, on the coupling
$\la$ and on the temperature $T$: $M(\Fc;m,\la,T)$.  The inverse function,
$\Fc(M;m,\la,T)$ is given explicitely by Eq.  \Ref{gapteq}:
$\Fc^{2}=(M^{2}+m^{2}-3\la\Delta(M))/3\la$. This function can be easily
plotted, see Figure \ref{figure1}, for real values of $M$. As seen, it is not
monotone and the inverse function, i.e., $M(\Fc;m,\la,T)$ is not unique. So the
gap equation has  two solutions in $M$ for each $\Fc$. In order to understand
that better we consider the graphic solution of Eq. \Ref{gapteq}.  For this we
write the equation in the form $M^{2}=r_{hs}(M)$ with the notation
$r_{hs}(M)={-m^{2}+3\la \Fc^{2}+3\la\Delta_{0}(M)}$ and plot the lhs. and the
rhs. of this equation as functions of $M^{2}$ in one picture, see figure
\ref{figure2}.  The curve for the rhs. starts at $M=0$ in
${r_{hs}}_{|_{M=0}}=-m^{2}\delta_{1}+3\la\Fc^{2}+\la T^{2}/4$ which follows
from \Ref{d0a} where we introduced the notation
\be\label{de1}
\delta_{1}=1-{3\la\over8\pi^{2}} \,.
\ee
The derivative is negative there (see the second term in the rhs. of Eq.
\Ref{d0a}) so that for $ -m^{2}\delta_{1}+3\la\Fc^{2}+\la T^{2}/4 > 0$ there
is a solution close to $M=0$. In the opposite case there is no real solution
there.  For large values of $M$, the curve describing the rhs. will be
dominated by the logarithm, see formula \Ref{d0}. For $3\la\ln
(M^{2}/2m^{2})/16\pi^{2} \approx 1$ it becomes larger than the lhs. and
delivers the other solution, seen in Figure 1 as the decreasing part of the
curve. It takes place for large values of $\la$ and $M$ (at any $T$) and is of
no relevance for physical applications.

For $-m^{2}\delta_{1}+3\la\Fc^{2}+\la T^{2}/4 = 0$ the curve of the rhs.
starts in the origin and the solution is exactly $M=0$. 

In general, the solution of the gap equation is non perturbative and can be
obtained only numerically. However, for the parameters in some range there are
two possibilities for approximative procedures.

\subsection{The Iterative Solution of the Gap Equation}
Equation \Ref{gapteq} can be iterated. For this we write it in the form
$M=\sqrt{r_{hs}(M)}$. Inserting some zeroth approximation $M^{0}$ into the
rhs.  the next approximation $M^{(1)}$ can be obtained which in turn must be
inserted into the rhs. and so on. This procedure converges until the
module of the derivative of the rhs. with respect to $M$ does not exceed
unity. This derivative takes its largest value at $M=0$. From this a
sufficient criterion for the convergence can be derived. Let us consider the
derivative
\[
{\pa r_{hs}\over \pa M }_{|_{{M=0}}}={-3\la T/ 4\pi\over
  2\sqrt{-m^{2}\delta_{1}+3\la\Fc^{2}+\la T^{2}/4}}\,.
\]
It is smaller than unit for 
\[
\la < {4m^{2}\over 3T^{2}}+{32\pi^{2}\Fc^{2}\over 3 T^{2}}+{8\pi^{2}\over 9}
+\sqrt{\left({4m^{2}\over 3T^{2}}+{32\pi^{2}\Fc^{2}\over 3
      T^{2}}+{8\pi^{2}\over 9}\right)^{2} -{m^{2}\over T^{2}}} \,.
\]
Also, even for $\la$ obeying this restriction, the denominator may become
small and make the iteration procedure diverge. In fact, this happens for $M$
close to zero. 

In case the sufficient conditions are satisfied the iteration procedure works
well.  It does so even in the complex region, i.e., for $M^{2}<0$ and down to
$T\rsim 0$. The solution appears in powers of $\sqrt{\la}$.

\subsection{The Approximative Solution}
\begin{figure}[!htbp]\unitlength=1cm
\begin{picture}(15,10)
\put(0,0){\epsfxsize=12cm
\epsffile{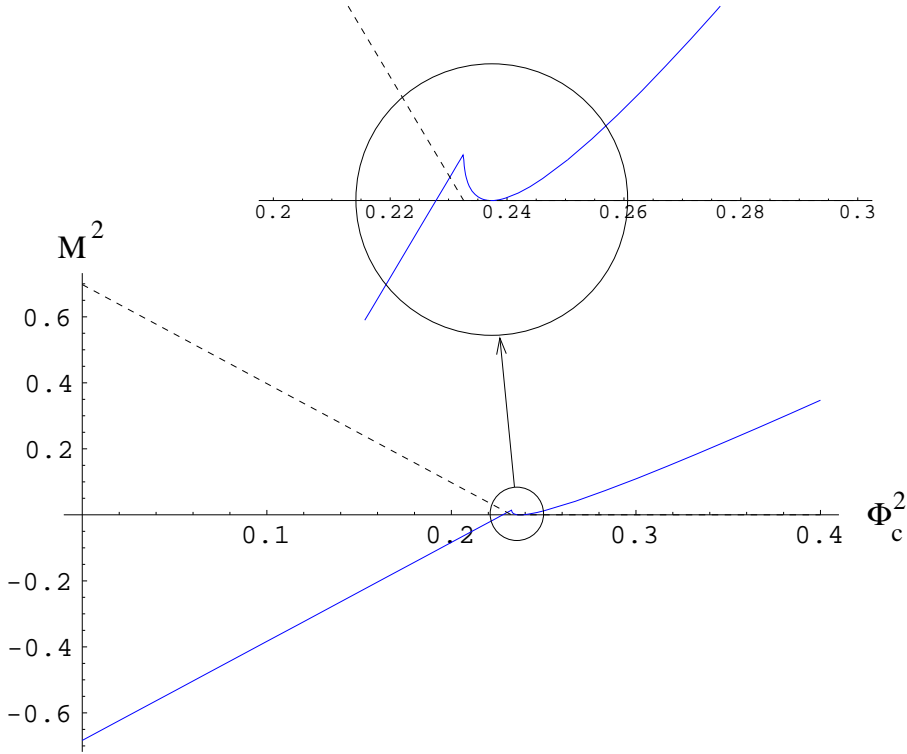}}
\end{picture}
\caption{Real part (solid line) and imaginary part (broken line) of the
  effective mass
  squared $M^{2}$ as function of $\Fc^{2}$ solving the gap equation for
  $\la=T=1$, the magnification shows details near $M=0$}
\label{figure3}
\end{figure}

Instead of iterating the Eq.  \Ref{gapteq} one may approximate it and solve
the approximated equation.  We restrict ourselves to temperatures of order
$T\rsim\la^{-\frac12}$. This is just the order where the phase transition
happens (see below). Then in the gap equation the approximation \Ref{d0a} may
be used.  So the approximated gap equation reads
\be\label{gapaeq}
M^{2}=-m^{2}+3\la\Fc^{2}+3\la\left\{{T^{2}\over 12}-{MT\over 4\pi}+{1\over
  16\pi^{2}}\left[M^{2}\left(\ln{\left(4\pi T\right)^{2}\over
      2m^{2}}-2\gamma\right)+2m^{2}\right]\right\}\,.
\ee
With the notation 
$\delta_{2}=1-\frac{3\la}{8\pi^{2}}\left[\ln{(4\pi T^{2})^{2}\over
    2m^{2}}-2\gamma\right]$ 
the solution reads
\be\label{app1}
M=-{3\la T\over 8\pi\delta_{2}}
+ \delta_{2}^{-\frac12}\sqrt{\left({3\la T\over
      8\pi \sqrt{\delta_{2}}}\right)^{2}-m^{2}\delta_{1}+3\la\Fc^{2}
+{\la   T^{2}\over 4}} \,.
\ee
Again, it can be seen that it is real for $T^{2}\ge
\left({4m^{2}\over\la}\delta_{1}-12\Fc^{2}\right)/\left(1+{9\la\over
    2\pi\delta_{2}}\right)$.  Thereby, for $T^{2}<{4m^{2}\over\la}\delta_{1}$
this is the case for $\Fc^{2}>\frac{1}{12}({4m^{2}\over\la}-T^{2})$ whereas
for $T^{2}>{4m^{2}\over\la}\delta_{1}$ the solution is real for any $\Fc$.
For $-m^{2}\delta_{1}+3\la\Fc^{2} +{\la T^{2}/4}<0$ the solution becomes
complex and the mass squared becomes negative.  The solution $M^{2}(\Fc^{2})$
of the gap equation is shown in figure \ref{figure3} for a temperature where
it has an imaginary part for small $\Fc$ and is real for larger $\Fc$.  It is
a nearly linear function and follows mostly the tree solution, $M=\mu$,
$\mu^{2}={-m^{2}+3\la\Fc^{2}}$. Only near $M=0$ it shows a nontrivial behavior
which is the non perturbative region for small $M$ discussed above.

The validity of the approximation leading to Eq. \Ref{gapaeq} can be
established as follows. First, consider large $T$. The solution is
$M=\sqrt{\la}T/2(1+O(\sqrt{\la}))$. Second, consider $T\sim 1/\sqrt{\la}$. The
solution is $M\lsim\sqrt{\la}T$. So in both cases the contributions neglected
in \Ref{d0a} are of order $\frac{M}{T}\sim\sqrt{\la}$ or smaller (for example,
at $-m^{2}\delta_{1}+3\la\fc^{2}+\la T^{2}/4=0$ the solution $M=0$ is exact).
Hence including higher order terms into \Ref{d0a} would result in corrections
of order $\sqrt{\la}$. On the other hand side dropping any of the terms in
\Ref{d0a} changes the solution quantitatively. It can be checked that in doing
so even the order of the phase transition changes.

\section{The Shape of the Effective Potential and its Extrema}\label{Sec4}
\begin{figure}[!htbp]\unitlength=1cm
\begin{picture}(10,7)
\put(0,-0.1){
\epsffile{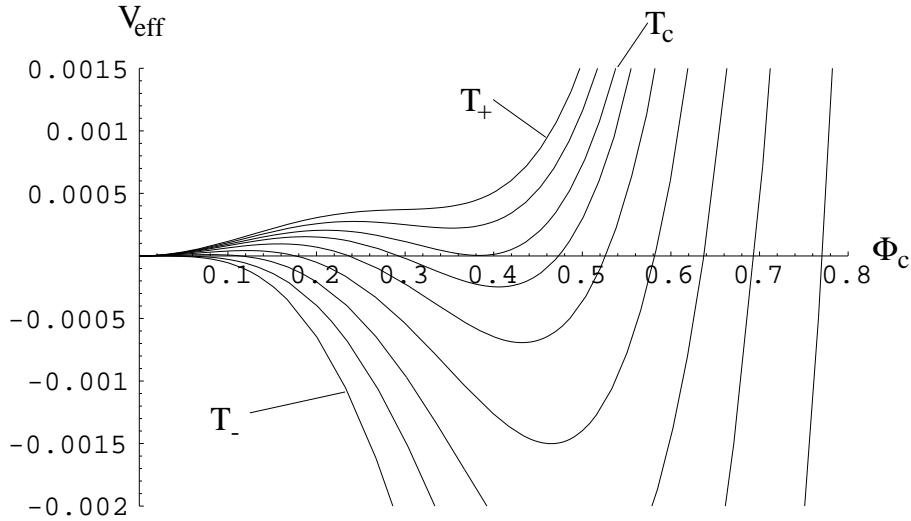}}
\end{picture}
\caption{The effective potential as function of $\fc$ for several temperatures
  in the region $T_{-}<T<T_{+}$ for $\la=1$}
\label{figure4}
\end{figure}

The effective mass and the effective potential can be calculated numerically
and plotted. In the temperature region where the phase transition takes place
the approximative solution \Ref{app1} of the gap equation together with
formula \Ref{V1a} makes it very easy to plot the effective potential as a
function of $\Fc$, see figure \ref{figure4}. It shows the typical behavior of
a first order phase transition. This important property can be shown exactly.

We investigate the extrema of the effective potential. Differentiating $V_{\rm
  eff}$ as given by Eq. \Ref{Veff2} we obtain 
\be\label{eabl}
{\pa V_{\rm eff}\over \pa \Fc^{2}}=
\left(-{M^{2}+m^{2}\over 6\la}+\frac12 \Fc^{2}+\frac12\Delta_{0}(M)\right)
{\pa M^{2}\over \pa\fc^{2}}+\frac12 M^{2}-\la\fc^{2} \,,
\ee
where \Ref{zh} was used. By means of the gap equation \Ref{gapteq} this
expression simplifies and from the condition $\pa V_{\rm eff}/\pa \fc^{2}=0$
we obtain
\be\label{fcex}
\la\Fc^{2}={M^{2}\over 2} \,.
\ee
This relation must be considered together with the gap equation where we
eliminate $\la\Phi_{c}^{2}$ by means of Eq. \Ref{gapteq} and obtain
\be\label{gapex}
M^{2}=2m^{2}-6\la\Delta_{0}(M)\,
\ee
as the gap equation in the extrema of the effective potential.

\begin{figure}[!htbp]\unitlength=1cm
\begin{picture}(10,7)
\put(0,0){\epsfxsize=12cm
\epsffile{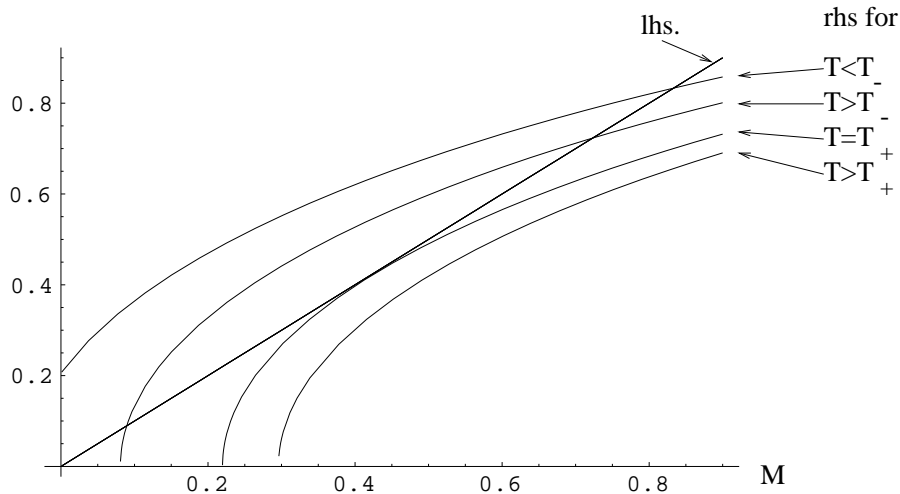}}
\end{picture}
\caption{Plot of the rhs. and of the lhs. ($=M$) of the gap equation in the
  minimum as function of $M$ for $\la=1$ and several temperatures}
\label{figuremiax}
\end{figure}

Eq. \Ref{gapex} has a form similar to the gap equation \Ref{gapteq}.
Again, we consider the graphical solution written in the form
$M=\sqrt{2m^{2}-6\la\Delta_{0}(M)}$.  Typical pictures are shown in figure
\ref{figuremiax}.

The curve for the rhs starts in $2m^{2}\delta_{1}-\la T^{2}/2$ with a positive
derivative (cf. Eq. \Ref{d0a}).  Therefore, for sufficiently small $T$ there
is one real solution. This is the position of a minimum of the effective
potential (for $T=0$ one obtains the known picture of the tree approximation).
For $T^{2}\ge 4m^{2}\delta_{1}/\la$ the starting point of the curve
representing the rhs. is below the origin and there are two real solutions.
Because the effective potential is a continuous function, this second solution
must be a maximum. It appears at smaller values of $\Fc$ than the minimum. So
\be\label{T-}
T_{-}={2m\over\sqrt{\la}}\sqrt{\delta_{1}}
\ee
is the lower spinodal temperature.  Raising the temperature lowers the curve
representing the rhs. further. The two solutions become closer to each other
until they merge finally. At that temperature the minimum and the maximum are
at the same place. Hence this is the upper spinodal temperature $T_{+}$. For
even higher temperature there is no real solution and the effective potential
is a monotone function -- the symmetry is restored.

Because these are exact, non perturbative properties of Eq. \Ref{gapex} it is
shown exactly that the phase transition is of first order.

It is important to note that the effective action is real in both its minima.
For $T<T_{-}$ there is only one minimum (that at finite $\Fc$). There the
effective potential is real. This is clear because the effectve mass is real
as shown just above and because the functions $\Delta_{0}(M)$ and $V_{1}(M)$
are real for real arguments. In $\Fc = 0$ the effective potentiual has a
maximum and it has an imaginary part. This corresponds to the instability of
this state.

Consider $T_{-}<T<T_{+}$. Here the effective potential has two minima, one in
$\Fc =0$ and the other one at finite $\Fc$. In both minima (as, in general,
for any $T>T_{-}$) the effective potetential is real. So in the given
approximation they are both stable. In fact, the effective potential should
have an imaginary part in the minimum at $\Fc =0$ for $T<T_{c}$ because there
is a finite probabiity for a tunneling transition to the lower lying minimum.
But this is beyond the given approximation.

It is interesting to note that in $T=T_{-}$ the effective mass at $\Fc =0$
vanishes. Also the second derivative is zero there, see Eq. \Ref{zabl}. So
this point is metastable as it must be in a first order phase transition.

These features can be seen explicitely in the expansion of Eq. \Ref{gapex} for
small $\la$. Using \Ref{d0a} in the rhs. of Eq. \Ref{gapex} we obtain
\be\label{Me}
M^{2}=2m^{2}-6\la\left\{{T^{2}\over 12}-{MT\over 4\pi}+{1\over
    16\pi^{2}}\left[M^{2}\left(\ln{(4\pi T)^{2}\over
        2m^{2}}\right)+2m^{2}\right]\right\} \,,
\ee
which can be easily resolved with respect to $M$. With the notations for
$\delta_{1}$  introduced in the preceding section and
$\delta_{3}=1+{3\la\over 8\pi^{2}}\left(\ln{\left(4\pi T\right)^{2}\over
    2m^{2}}-2\gamma\right)$ we obtain
\be\label{Miax}
M_{\rm Min/Max}={3\la T\over 4\pi \delta_{3}}\pm
\sqrt{
\left({3\la T\over 4\pi \delta_{3}}\right)^{2}+{2m^{2}\delta_{1}\over
  \delta_{3}}-{\la T^{2}\over 2\delta_{3}} \,,
}\ee
where the upper sign corresponds to the minimum and the lower sign to the
maximum of the effective potential. The condensate $\Fc$ is related to this
mass by Eq. \Ref{fcex}. The lower spinodal temperature $T_{-}$ is  the
temperature at which the maximum appears at $M_{\rm Max}=0$ when raising the
temperature.  It is the same as given by formula \Ref{T-}. Also, the upper
spinodal temperature $T_{+}$ follows from formula \Ref{Me}. It is the
temperature for which the positions of the maximum and of the minimum
coincide, i.e., where the square root vanishes:
\be\label{T+}
T_{+}= {2m\over\sqrt{\la}}\sqrt{{\delta_{1}\over 1-{9\la\over
      8\pi^{2}\delta_{3}}}} \,.
\ee
Within the given precision\footnote{bearing in mind that corrections of order
  $\la^{2}$ and higher follow from the graphs contained in $W^{\rm 1PI}(0)$,
  Eq. \Ref{Vefft}}, i.e., in the lowest nontrivial order in $\la$,
these temperatures can be written as
\be\label{T+-}
 T_{-}={2m\over\sqrt{\la}}\left(1-{3\la\over 16\pi^{2}}\right),
\qquad T_{+}={2m\over\sqrt{\la}}\left(1+2{3\la\over 16\pi^{2}}\right)\,.  
\ee
They are both of order $m/\sqrt{\la}$ justifying the approximation made in the
gap equation. The interval between $T_{-}$ and $T_{+}$ is of order
$\sqrt{\la}m$.

\begin{figure}[!htbp]\unitlength=1cm
\begin{picture}(10,7)
\put(-3,-18.3){
\epsffile{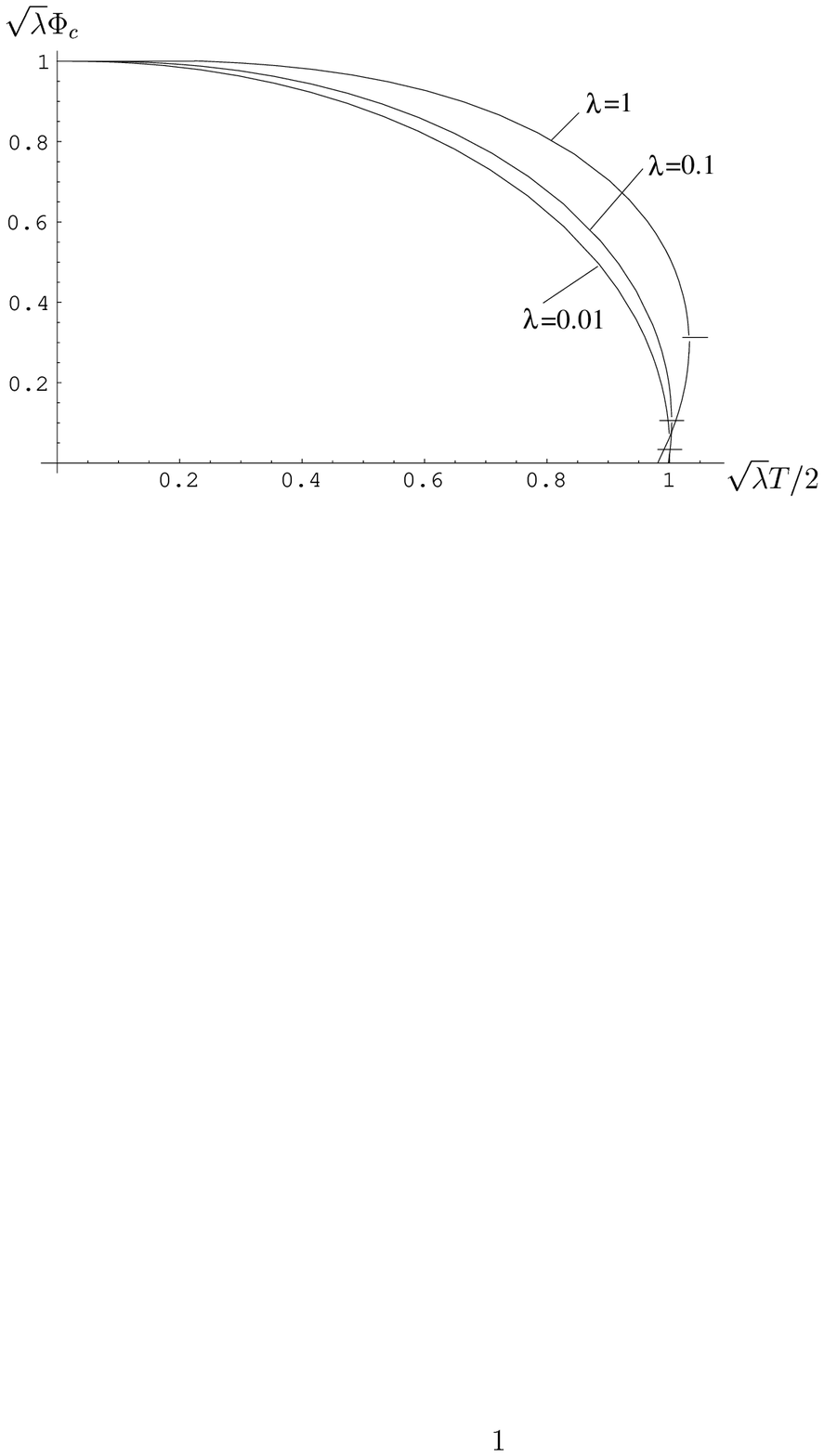}}
\end{picture}
\caption{$\sqrt{\la}\fc=M /\sqrt{2}$ in the minimum (resp. maximum) (upper resp. lower
  part of the curves) of the effective potential as function of the
  $T\sqrt{\la}/2$ for several values of the coupling $\la$}
\label{figure5}
\end{figure}

The critical temperature $T_{c}$, which is the temperature where the effective
potential in its minimum at finite $\fc$ equals
its value at $\Fc=0$, is in between these two temperatures, closer to $T_{+}$.

The positions of the minimum and the maximum of the effective potential as
function of the temperature can be calculated from Eqs. \Ref{gapex} and
\Ref{fcex} numerically. Also, they can be taken from \Ref{Miax} except for
small values of $T$ where the approximation made in \Ref{Me} is not valid
(there the solution $M$of Eq. \Ref{gapex} has a nonanalytic behaviour of the
type $M\raisebox{-5pt}{$\stackrel{\sim}{\scriptstyle{T\to 0}}$}\sqrt{2}m+O(\exp(-{\sqrt{2}m\over T})) $).
The result is shown in figure \ref{figure5} for several values of the coupling
$\la$.

It is possible to calculate in a quite easy way the depth of the minimum at
$T=T_{-}$. For this we need the potential at $\Fc=0$ for this temperature.
There the gap equation has the (exact) solution $M=0$. This is just the
situation when the curve representing the rhs. in figure \ref{figure2} starts
from zero.  Than by means of \Ref{V1a} and \Ref{Vefft} we obtain
\[
{V_{\rm eff}}_{|_{\Fc=0, \ T=T_{-}}}=
-\frac{m^{4}}{12\la}-{\pi^{2}T_{-}^{4}\over 90} \,.
\]

In order to calculate the effective potential in its minimum at $T=T_{-}$ we
use \Ref{Veff2} and insert $\fc$ from Eq. \Ref{fcex}. We obtain
\[
{V_{\rm eff}}_{_{|Min, \ T=T_{-}}}= {M_{\rm Min}^{4}\over 24\la}-{{m^{2}M_{\rm
      Min}^{2}\over 6\la}-{m^{4}\over 12\la}+V_{1}(M_{\rm Min})} \,. 
\]
Now we insert $T_{-}$ from Eq. \Ref{T+-} and $M_{\rm Min}$ from \Ref{Miax}.
Expanding $V_{1}$ according to \Ref{V1a} we obtain for the difference
\be\label{deepth} {V_{\rm eff}}_{|{Min, \ T=T_{-}}} - {V_{\rm eff}}_{|_{\Fc=0,
    \ T=T_{-}}}=-{9\over 16}{\la m^{4}\over \pi^{4}} \,, \ee
where higher corrections in $\la$ are dropped. This is the depth of the
minimum at the lower spinodal temperature. 

\begin{figure}[!htbp]\unitlength=1cm
\begin{picture}(15,5)
\put(0,0){\epsfxsize=10cm
\epsffile{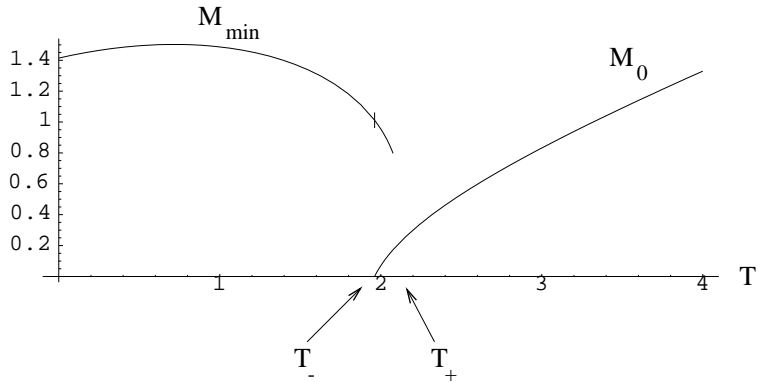}}
\end{picture}
\caption{The Debye mass, i.e., the effective mass in the positions of the
  minima of the effective potential as function of the temperature for $\la=1$}
\label{figure6}
\end{figure}

It is interesting to calculate the effective mass $M$ in the minima of the
effective potential as function of the temperature. This is the so called
Debye or temperature dependent mass. For $\Fc=0$ it follows from Eq.
\Ref{app1}. At $T=T_{+}$ we note the value
$M_{0}=\frac34(\sqrt{3}-1)\sqrt{\la}m$. In the other minimum it is what we
denoted by $M_{\rm Min}$, given by Eq.  \Ref{Miax} (upper sign). At $T=0$
we note $M_{\rm Min}=2m^{2}$ which is the tree value. At $T=T_{-}$ we have
$M_{\rm Min}=3\sqrt{\la}m/2\pi$ and at $T=T_{+}$ we have $M_{\rm
  Min}=3\sqrt{\la}m/4\pi$.  Both are shown in figure \ref{figure6} as function
of the temperature.

Another important check of the consistency of the given approximation is that
in the minimum of the effective potential the term linear in $\eta$ in the
effective action \Ref{gat} vanishes exactly. This can be seen by inserting
$\Fc$ from \Ref{fcex} and using Eq. \Ref{gapex}. Note that from this the
disappearance of all tadpole diagrams in the Green functions, not only in the
effective action, follows. 

Also, the connection between the effective mass and the second derivative of
the effective potential in the minimum can be checked. For this reason we
calculate $\pa^{2}V_{\rm eff}/\pa (\fc^{2})^{2}$ from \Ref{eabl} and obtain by
means of \Ref{fcex} and \Ref{gapex}
\be\label{zabl} 
{\pa^{2}V_{\rm eff}\over\pa (\fc)^{2}}={1+3\la\Delta_{0}'(M)\over 
1-3\la\Delta_{0}'(M)}M^{2} \,,
\ee
with $\Delta'=(\pa \Delta /\pa M^{2})$ which is the expected relation within
the given approximation.

\section{Comparison with different approaches}\label{Sec5}
Since the pioneering papers back to the 70th there have been several steps in
the developement of the topic. Already in \cite{DolanJackiw} the necessity of
summing 'daisy' and 'super daisy' diagrams was underlined. Perhaps the
first summation of 'daisy' diagrams (also called 'ring' diagrams) for different
models of quantum field theories was carried out
 in \cite{Takahashi:1985vx}, later on it was reconsidered in 
\cite{Carrington:1992hz} with specific application to the electroweak phase
transition in the Standard Model. In the latter paper 
 using  some high temperature approximations a first order
phase transition was obtained where also the imaginary part comes out right.
As we will see below, this is a qualitatively good approximation. Detailed 
information about properties of the high temperature  daisy approximation 
can also be found in \cite{LindeLinde}.

\subsection{One Loop Approximation}
The simplest approximation is the 'pure Tr ln', which means the one loop
contribution to the effective action without summing up any 'daisy'
diagrams. The corresponding expression is simply
\be\label{Va}
V_{\rm eff}^{\rm a}=-\frac{m^{2}}{2}\fc^{2}+\frac{\la}{4}\fc^{4}+V_{1}(\mu^{2})
\ee
with $\mu^{2}=-m^{2}+3\la\fc^{2}$ and $V_{1}(m)=\frac12 \Tr \ln
(p^{2}+m^{2})$. The function $V_{1}$ is given in the Appendix,
\Ref{A1}, \Ref{V1}. The behaviour of this effective potential is in general 
well known, but there are some subtles. In order to encounter them we consider
the extrema of \Ref{Va} by equating zero its derivative
\be\label{anull}
{\pa V_{\rm eff}^{\rm a}\over \pa
  \fc^{2}}=-\frac{m^{2}}{2}+\frac{\la}{2}\fc^{2}+\frac32 \la\Delta_{0}(\mu^{2})\,,
\ee
whereEq. \Ref{zh} had been used. This can be rewritten in the form
\be\label{gapa}
\mu^{2}=2m^{2}-9\la\Delta_{0}(\mu)\,.
\ee
\begin{figure}[!htbp]\unitlength=1cm
\begin{picture}(15,5.5)
\put(0,0){\epsfxsize=9cm
\epsffile{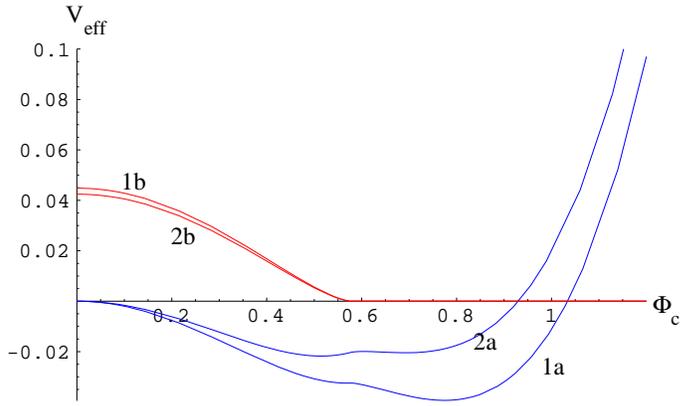}}
\end{picture}
\caption{The effective potential in one loop approximation for $\la=1$. Curve
  1a resp. 1b shows the real resp. imaginary part for a temperature where the
  effective potential is real in the lower minimum, curves 2a resp. 2b are for
  a higher temperature and the effective potential is complex in the lower
  minimum.} 
\label{figureVa}
\end{figure}
This relation is like  Eq. \Ref{gapex} with the difference that the effective
mass is $\mu^{2}=-m^{2}+3\la\fc^{2}$ in this case. In complete analogy to Eq.
\Ref{gapex} the solution can be investigated graphically. Figure
\ref{figuremiax} applies changing merely the coupling according to
$\la\to\frac32 \la$. Hence the effective potential has a minimum for
$T^{2}<8m^{2}/3\la$. For $T^{2}\ge 8m^{2}/3\la$ there are both, a minimum and
a maximum. Raising $T $ further they merge and disappear. These are properties
of real solutions, i.e., for $\mu^{2}>0$, i.e., $\la\Fc^{2}>m^{2}/3$. For
$\la\Fc^{2}<m^{2}/3$ the effective potential is complex. Therefore, there might
be another minimum if considering only the real part. The effective potential
resp. its real and imaginary parts can be easily plotted using formula
\Ref{V1} and the integral representation in \Ref{S2}, see figure
\ref{figureVa}. 
\begin{figure}[!htbp]\unitlength=1cm
\begin{picture}(15,5.5)
\put(0,0){\epsfxsize=11cm
\epsffile{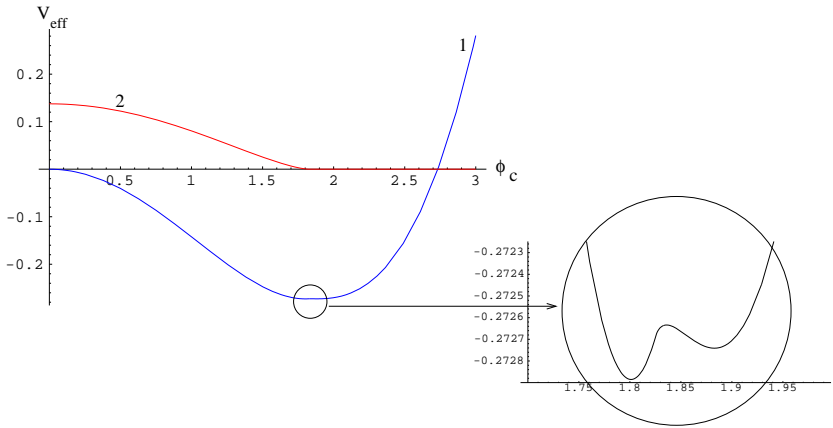}}
\end{picture}
\caption{The effective potential in one loop approximation for $\la=0.1$. 
  Curves 1 resp. 2 show the real resp. imaginary part, the magnification shows
  both minima present.}
\label{figure1a}
\end{figure}

The general behaviour is that of a second order phase transition. In a
narrow temperature region $T^{2}\rsim 8m^{2}/3\la$ the real part has
two minima. At $T^{2}<8m^{2}/3\la$ the left one is lower and for
$T^{2}>8m^{2}/3\la$ the right one is the lower one. In the left one
the effective potential is complex and in the right one it is real. It
seems that the very tiny dip which is shown in figure \ref{figureVa}
for $\la=1$ and in figure \ref{figure1a} for $\la=0.1$ in the
magnification had not been observed before. It is clear that the pure
one loop effective potential is unphysical because it is complex in
the minimum at some temperatures.  The position if the extrema are
shown in figure \ref{figver}, curves 3,4 and 5 as functions of
$T$. Curve 3 shows the minimum where $V_{\rm eff}^{\rm a}$ is real, on
curve 5 it is complex and curve 4 shows the position of the maximum.

\subsection{'Simple Daisy' Diagrams Summed Up}
After summing up all 'daisy' and 'super daisy' diagrams, which resulted in the
effective potential \Ref{Vefft} and the gap equation \Ref{gapteq} it is quite
simple to go the step back to the simple 'daisy' diagrams summed up by
substituting the tree value $\mu^{2}=-m^{2}+3\la\Fc^{2}$ instead of the
effective mass $M^{2} $ at two places in these formulae which results in
\be\label{Vb}
V_{\rm eff}^{\rm b}=-\frac{m^{2}}{2}\fc^{2}+\frac{\la}{4}\fc^{4}-\frac34\la\Delta_{0}(\mu)+V_{1}(M)
\ee
with
\be\label{Mb}
M^{2}=-m^{2}+3\la\fc^{2}+3\la\Delta_{0}(\mu) \,,
\ee
where $V_{1}$ resp. $\Delta_{0}$ are the same functions \Ref{A1} resp.
\Ref{d0} as in section \ref{Sec2}. However, we have to show that \Ref{Vb} and
\Ref{Mb} is just the result of summing the simple 'daisy' diagrams. For this
we have to go back to formula \Ref{Gammapert} and must pick up the daisy
diagrams from $W^{\rm 1PI}$. We obtain
\be\label{Vb1}
V_{\rm eff}^{\rm b}=-\frac{m^{2}}{2}\fc^{2}+\frac{\la}{4}\fc^{4}+
\frac12
\Tr\ln (p^{2}+m^{2})
-\frac18 \ \raisebox{-10pt}{\epsffile{eight.eps}}
-\frac{1}{16} \ \raisebox{-10pt}{\epsffile{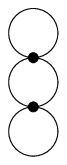}}
-\frac{1}{48} \ \raisebox{-10pt}{\epsffile{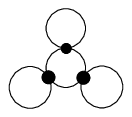}}-\dots \,.
\ee
To be explicit, here we used the notation $\Tr
=\frac{1}{\beta}\sum_{l=-\infty}^{\infty}\int {\d \vec{k}\over (2\pi)^{3}}$
and $\Tr \ln (p^{2}+m^{2}) =-\Tr\Delta = -V_{1}(m)$ where
$\Delta=1/(p^{2}+m^{2})$ is the line in the graphs (in momentum
representation). The vertex factors are $-6\la$. Now, the analytic expression
for the graphs reads
\bea\label{Vb2}
V_{\rm eff}^{\rm b}&=&-\frac{m^{2}}{2}\fc^{2}+\frac{\la}{4}\fc^{4}  
+\frac12
\Tr\Big[\ln (p^{2}+m^{2})  \\
&& \qquad
-\frac14 \ {-6\la\Delta_{0}(m)\over p^{2}+m^{2}}
-\frac18 \ {(-6\la\Delta_{0}(m))^{2}\over (p^{2}+m^{2})^{2}}
-\frac{1}{24} \ {(-6\la\Delta_{0}(m))^{3}\over (p^{2}+m^{2})^{3}}
-\dots
\Big] \nn \,,
\eea
where $\Delta_{0}(m)=\Tr 1/(p^{2}+m^{2})$ is in fact the same as \Ref{delta0}
resp. \Ref{D0} written for $T\ne 0$ and with $m$ instead of $M$. The graph
'eight', \raisebox{-10pt}{\epsffile{eight.eps}} , can be represented in two
ways
\[ \raisebox{-10pt}{\epsffile{eight.eps}}=-6\la \left(\Tr{1\over
  p^{2}+m^{2}}\right)^{2}=-6\la \ \Tr {\Delta_{0}(m)\over p^{2}+m^{2}}
\]
(cf. Eq. \Ref{D0}). Eq. \Ref{Vb2} is already the expansion of the logarithm
($\ln (1+\epsilon)=\epsilon-\epsilon^{2}/2+\epsilon^{3}/3-\dots$) except for
the 'eight', which enters with the wrong sign. In writing the coefficient in
front as $\frac 12=-\frac12+1$ (times $3\la$ after cancelling a factor of 2) we
obtain
\beao
V_{\rm eff}^{\rm b}&=&-\frac{m^{2}}{2}\fc^{2}+\frac{\la}{4}\fc^{4}  \\
&&\qquad +\frac12 \ \Tr \Bigg[ \ln (p^{2}+m^{2})-\frac12 \ 
{3\la\Delta_{0}(m)\over
  p^{2}+m^{2}}+\ln \left(1+{3\la\Delta_{0}(m)\over p^{2}+m^{2}}\right)\Bigg] \\
&=&-\frac{m^{2}}{2}\fc^{2}+\frac{\la}{4}\fc^{4}
-\frac34\la\Delta_{0}^{2}(m)+\frac12\Tr\ln
\Big(p^{2}+m^{2}+3\la\Delta_{0}(m)\Big)\,.
 \eeao
 This is just \Ref{Vb} and \Ref{Mb}.
 
\begin{figure}[!htbp]\unitlength=1cm
\begin{picture}(15,5.5)
\put(0,0){\epsfxsize=9cm
\epsffile{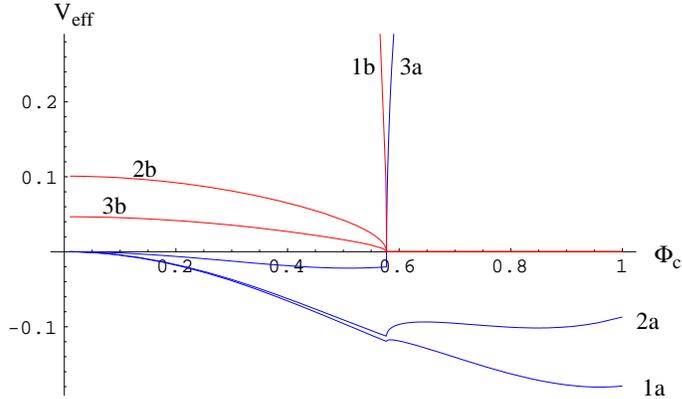}}
\end{picture}
\caption{The effective potential with 'simple daisy' graphs summed up for
  $\la=1$. Curves 1a, 2a, 3a resp. 1b, 2b, 3b show the real resp. imaginary
  parts for $T=1, \ 1,4, \ 3.4$.} 
\label{figured}
\end{figure}

\begin{figure}[!htbp]\unitlength=1cm
\begin{picture}(15,7.5)
\put(0,0){\epsfxsize=11cm
\epsffile{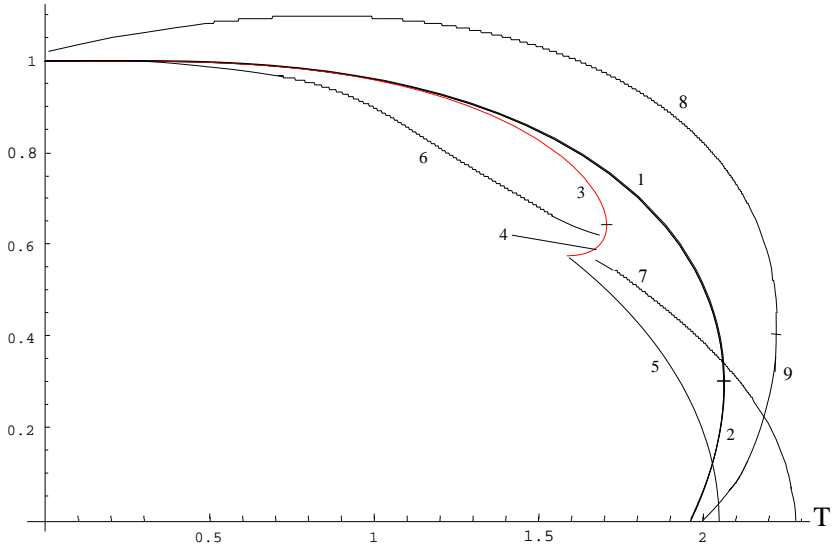}}
\end{picture}
\caption{Positions of minima and maxima of the effective potential at $\la=1$ 
in  different approximations, see text}
\label{figver}
\end{figure}
The investigation of the extrema of $V_{\rm eff}^{\rm b}$ is more complicated
that in the other cases and not very interesting because this case in anyway
not physical. Nevertheless it is interesting to comment on them. $V_{\rm
  eff}^{\rm b}$ can be plotted qute easily using the above formulae. It is
shown in figure \ref{figured} as function of $\fc$ for several values of the
temperature. It has an imaginary part for $\la\fc^{2}<m^{2}/3$. At
$\la\fc^{2}=m^{2}/3$ it has a cusp and above it is real. Again, the real part
has two minima in a narrow temperature range. In general, the behaviour is
quite similar to that of the 'simple Tr ln'. The positions of the minima can
be found numerically and they are shown in figure \ref{figver} as curves 6 and
7 (the maximum is not shown there). In the minimum represented by curve 7
resp. 6 the effective potential is complex resp. real.

\subsection{High Temperature Approximation and higher loop}
Frequently high temperature approximations for the functions $V_{1}$ and
$\Delta_{0}$ are used which are given by parts of the Eqs. \Ref{V1a},
\Ref{d0a}. Another type of approximations used can be obtained by taking only
the $l=0$ contribution in the sum over the Matsubara frequencies. A typical
example is given in \cite{Takahashi:1985vx}. There, the expression
\be\label{Vtak}
V_{\rm eff}^{\rm t}=-\frac{m^{2}}{2}\fc^{2}+\frac14 \la\fc^{4}
+\frac{T^{2}}{24}(-m^{2}+3\la\fc^{2})-\frac{T^{2}}{12\pi}\left({\la T^{2}\over
    4}-m^{2}+3\la \fc^{2}\right)^{\frac32}
\ee
has been obtained (see also \cite{Carrington:1992hz}). Here, the phase
transition is first order, the effective potential is real in its minima. The
positions of the maximum are shown in figure \ref{figver} ( curve 8 resp.
curve 9). It is seen, the behaviour is similar to that of the correct
solution (curves 1 and 2), but the numbers are about of 10 ... 20 \% higher.
So we conclude that the simple formula \Ref{Vtak} works basically well and
gives a result which is qualitatively correct and quite close to the right
numbers.

In the paper \cite{Espinosa:1992gq} the daisy and super daisy graphs have been
summed up by direct summation of the graphs performing the difficult task of
calculating the combinatorical coefficients (which in the present paper was
done implicitely in section \ref{Sec2}). A first order transition resulted.
Than the next graph, \epsfxsize=1cm\raisebox{-3pt}{\epsffile{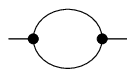}} \ ,
had been included in the resummation. However, the combinatorics could not
been done to the end. This is probably the reason that an unwanted imaginary
part appeared which led the authors to the conclusion that the subleading
terms they calculated are not consistent.

In another paper \cite{Nachbagauer:1995ur} a similar approach had been
undertaken trying to solve the gap equation with two graphs.  In a subsequent
unpublished paper \cite{Nachbagauer:1994aj} also, a first order transition was
proposed.

\section{Conclusions}

In the foregoing sections the effective potential for a scalar theory with
spontaneous symmetry breaking and finite temperature is calculated. All
'daisy' and 'super daisy' diagrams are summed up and the properties of the
resulting gap equation are investigated. It is shown exactly that the
phase transition is first order. The imaginary part of the effective action
comes out in the correct way. 

It is important to notice that higher loop corrections cannot change these
features. The point is that they are all expressible in terms of Feynman
graphs with propagators $1/(k^{2}+M^{2})$ (in momentum space) where the mass
$M$ is solution of the gap equation. As shown it is positive in the physical
region, in the minima of the effective potential for instance (see figure
\ref{figure6}). Consequently all graphs have lines with non vanishing
denominator. As the theory is Euclidean, no more infrared divergencies can
occur. A special discussion deserves the contribution of the graph
\epsfxsize=1cm\raisebox{-3pt}{\epsffile{ryba.eps}} . As the vertex factor is
$\la\fc$ and in the minimum of the effective potential we have $\sqrt{\la}\fc
\sim 1$ it seems to contribute the same order as the daisy graph. However,
this cannot change the conclusion drawn above. It is only the quantitative
characteristics of the phase transition which may become altered, not the
qualitative ones like its order.

Natural extensions of the present work is the inclusion of fermions, gauge
bosons and external fields. 

\section*{Acknowledgement}

VS thanks DFG for support from the grant 436 UKR 17/24/98 and the
University of Leipzig for kind hospitality.

\section*{Appendix}\label{SecA1}
The function $V_{1}(M)$ appearing in \Ref{Vefft} is the notation for the
'Tr ln' term:
\be\label{A1} V_{1}(M)= - \frac12 \Tr \ln \Delta(M)
=\frac{1}{\beta}\sum_{l=-\infty}^{\infty}\int{\d \vec{k}\over (2\pi)^{3}} \ln
\left(\omega_{l}^{2}+\vec{k}^{2}+M^{2}\right) \ee
with $\beta=1/kT$ and $\omega_{l}=2\pi l/\beta$. 
Removing the \uv divergencies and taking into account the normalization
conditions the explicit expression reads
\be\label{V1}
V_{1}(M)={1\over 64\pi^{2}}\left\{4m^{2}M^{2}+ M^{4}\left[\ln{M^{2}\over
      2m^{2}} -\frac32\right] \right\}
-{M^{2}T^{2}\over2\pi^{2}} \, S_{2}\left({M\over T}\right)\,,
\ee
where the last term is the temperature dependent contribution.  The function
$S_{2}$ can be represented as a fast converging sum over the modified Bessel
function $K_{2}$ and by an integral representation which is suited for complex
values of the argument as well:
\be\label{S2} S_{2}(x)=\sum_{n=1}^{\infty}\frac1{n^{2}} \ K_{2}(nx)={1\over
  3x^{2}} \ \int_{x}^{\infty}\ \d n \ {\left(n^{2}-x^{2}\right)^{3/2}\over
  \e^{n}-1} \,.  \ee
Its expansion for small $x$ reads
\be\label{S2a} S_{2}(x)={\pi^4\over 45 x^2}-{\pi^2\over 12}+{\pi x\over
  6}+{x^2\over 32}\left(2\gamma-\frac32+2\ln\frac{x}{4\pi}\right)+O(x^3)\,,
\ee
giving rise to the high temperature expansion of $V_{1}$ which is at once the
expansion for small $M$:
\bea\label{V1a} V_{1}(M)&=&
{-\pi^{2}T^{4}\over 90}+{M^{2}T^{2}\over 24}-{M^{3}T\over 12\pi}  \\
&&+{1\over 64\pi^{2}}\left\{M^{4}\left[\ln{\left(4\pi T\right)^{2}\over
      2m^{2}}-2\gamma \right]+4m^{2}M^{2}\right\}+M^{2}T^{2} \ O\left({M\over
    T}\right)\,. \nn \eea
Here $\gamma$ is the Euler constant.

Similar formulae hold for the 'daisy' graph. Although they can be derived by
means of \Ref{zh} it is useful to have them separately:
\bea\label{d0}
\Delta_{0}(M)&=&\frac{1}{\beta}\sum_{l=-\infty}^{\infty}\int{\d \vec{k}\over
  (2\pi)^{3}} \ {1\over
\omega_{l}^{2}+\vec{k}^{2}+M^{2}} \nn \\
&=&
{1\over 16\pi^{2}}\left\{ 2 m^{2}+M^{2}\left[\ln\left({M^{2}\over
        2m^{2}}\right)-1\right]\right\}+{M T\over
    2\pi^{2}}\, S_{1}\left({M\over T}\right) \,,
\eea
where, again, the last term is the temperature dependent contribution with
\be\label{S1}
S_{1}(x)=\sum_{n=1}^{\infty}\frac1n \ K_{1}(nx)={1\over x} \ \int_{x}^{\infty}\ \d n \
{\sqrt{n^{2}-x^{2}}\over \e^{n}-1} \,.
\ee
The corresponding expansions are
\be\label{S1a}
S_{1}(x)={\pi^{2}\over 6x}-{\pi \over 2}-{x\over
  8}\left(2\gamma-1+2\ln\frac{x}{4\pi}\right)+O(x^2)
\ee
and
\be\label{d0a}
\Delta_{0}(M)={T^{2}\over 12}-{MT\over 4\pi}+{1\over
  16\pi^{2}}\left\{M^{2}\left(\ln{\left(4\pi T\right)^{2}\over
      2m^{2}}-2\gamma\right)+2m^{2}\right\}+MT \ O\left({M\over T}\right)\,.
\ee

\bibliographystyle{unsrt}
\bibliography{ges1}
\end{document}